# Online tools for public engagement: case studies from Reykjavik

Iva Bojic[a], Giulia Marra[b] and Vera Naydenova[c]

[a] *SENSEable City Laboratory, Singapore-MIT Alliance for Research and Technology, 1 Create Way, Singapore, Singapore*
(ivabojic@mit.edu)
[b] *Politecnico di Torino, corso Duca degli Abruzzi 24, 10129 Torino, Italy*
(giulia.marra@polito.it)
[c] *Social Design – Arts as Urban Innovation, University of Applied Arts Vienna, Oskar Kokoschka-Platz 2, 1010 Vienna, Austria*
(vera.naydenova@student.uni-ak.ac.at)

## Introduction

With the ubiquity of Internet technologies and growing demands for transparency and open data policies, the role of social networking and online deliberation tools for public engagement in decision-making has increased substantially in the last decades. Across the globe, officials and non-governmental organizations respectively or jointly have introduced online platforms in order to involve the public at different stages of policy initiation and implementation and at various degrees of participation - from information to consultation to collaborative decisions (Habermas, 2001). The nature of such platforms is still experimental and their success varies; power to influence ultimate choices is hardly ever delegated to the citizens, the rhetoric of direct democracy and citizen empowerment often serving to disguise the factual lack of political and administrative will and dedication to public participation (Papacharissi, 2002). Partnerships with civic organizations could be key for more efficient management of time and resources, as well as better outreach and access to participation. However, experience in establishing such public-private partnerships is still lacking, and the implications for personal data protection and misuse of online media for political steering should be considered. What is more, attention should be focused on the cognitive limitations and psychological predispositions of those engaging in participation online, as research in recent years reveal a positive correlation between reported use of social media and levels of civic engagement, but the causality behind this has not been studied enough (Boulianne, 2015, Ruckner, 2016).

The aim of this paper is to assess the benefits, challenges and successful methods for public engagement through different online media using example of Reykjavik in Iceland. With Iceland being one of the most digitally connected countries where almost 98% of people have access to Internet in their homes (We are social, 2016), choosing online tools seems like a logical choice for involving public in participatory democracy. Moreover, Iceland makes a good case study as the ideas about new ways to connect people together to participate in democracy, politics and civic life were born there after its economic collapse in 2008. Since then Iceland have developed open source tools and methods to promote online, democratic debate and to increase citizens' participation in their community and worldwide. After the economic crisis, many citizen initiatives emerged in an attempt to tap the potential of digital platforms to increase access to information, transparency and accountability.

## Methodology

The research presented in this paper was developed within the framework of the international winter school *Co-Creating Urban Spaces: The Transformation of the Given City* organized by the COST Action *People Friendly Cities in a Data Rich World* (TU1204) in March 2016 in Iceland. Using Reykjavik as a site for in-situ exploration, participants - early career investigators, architects and entrepreneurs working in the area of urban development, social innovation, engineering and design - were given the opportunity to explore various local approaches to urban innovation and community engagement through lectures and field work on real case studies. In this paper we present the analysis of how social media are used by different public bodies to enhance public participation in deliberative democracy. We collected and reviewed published information on the subject and carried out a field base assessment, involving structured interviews with different government representatives and urban policymakers.

## Public e-Participation in Reykjavik

Although various options of interaction between public bodies and citizens' groups are offered in Reykjavik, in the paper we focus on the following four: (1) Reykjavik Facebook page; (2) citizens of Breiðholt Facebook closed group; (3) online forum Better Reykjavik; and (4) e-deliberation platform/consultation forum Better Neighborhoods.

### *Reykjavik Facebook Page*

This page started 3 years ago and today has more than 10 000 friends from all over the world. It serves as a forum to improve communication between Reykjavik and its citizens. Although the page has an Editorial Board composed of city officials who publish city news, it is not only a one-way communication as the Board provides answers to all comments and questions. However, questions posted on this page are not treated as formal requests, but rather this page serves as a sort of informal communication bridge: everyone gets a reply and city professionals always put their names under their replies, so that users know who answered their question or addressed their comment. The page is not used to consult citizens or to collect their ideas, but to promote offline citizens' participation.

### *Citizens of Breiðholt Facebook Closed Group*

The neighborhood of Breiðholt in Reykjavik has a citizen organization who is responsible for Facebook closed group started a couple of years ago with a few hundred members. Today it represents a communication pathway between the public bodies and their citizens with more than 5000 members where Breiðholt citizens can bring practical and fun ideas or suggestions that potentially can improve the quality of the neighborhood. When it was started, communication via the page was mostly negative (e.g. negative remarks about neighborhood shortcomings), but after introducing some rules, as well as people started noticing things are turning for better, the group turned into an active media for constructive comments and information flow on various neighborhood topics.

*Better Reykjavik*

Better Reykjavik is an online consultation forum where citizens are given the chance to present their ideas on different issues regarding their city. This effort is a result of an open collaboration between Reykjavik City Council and the Citizens Foundation, a non-profit organization based in Reykjavik, who created the platform. Today this platform enables citizens to voice, debate and prioritize ideas to improve their city, creating an open discourse between community members and city council and also giving the voters a direct influence on decision making. Forum is opened for anyone to read it, but only registered users can participate by presenting their ideas, viewing other ideas, arguing issues, voicing their opinion and by rating ideas. The best ideas chosen by the forum participants are then formally address by the city officials in the following manner: each month, five top rated ideas in all categories (i.e. tourism, operations, recreation and leisure, sports, human rights, art and culture, education, transportation, planning, administration, environment, welfare, various) with up to one top rated idea in each category are being presented in front of the appropriate committee.

*Better Neighborhoods*

Better Neighborhoods platform is based on ideas on participatory budgeting – promoting public participation in decision making beyond what is normally seen in a representative democracy. Citizens can submit their ideas on projects that they think will improve their neighborhoods and the city officials evaluate costs and feasibility of each project followed by citizens voting on the ideas. Each voter is empowered to decide how to distribute the total budget amount to projects that are relevant to him/her and this helps citizens to understand the realities of budgeting. The budget amount for this project has stayed the same during the last four years (i.e. 300 million) and in 2015 corresponded to 0.35% of the city total budget. The percentage of people who voted dropped from 8.1% in 2012 to 7.3% in 2015. The projects that can be proposed have to enhance the quality of the residents' surroundings, increase possibilities for recreation and social gatherings or opportunities for games and leisure, encourage cycling or improve conditions for pedestrians and public transportation users. Unlike in the case of Better Reykjavik, here ideas can be posted only once per year when the call for ideas is opened.

*Assessing citizens' satisfaction*

Although Better Reykjavik won the European award in the e-Democracy Awards in 2011 and Better Neighborhoods won Nordic Best Practice Challenge in the category Public Communication in 2014 (Iceland's Citizens Foundation 2016), the city officials wanted to heard their citizens' thoughts. Therefore, they asked Institute of Public Administration and Politics, University of Iceland to conduct online survey on a sample of 2500 citizens of Reykjavik with the purpose of assessing how Better Reykjavik and Better Neighborhoods are perceived in public and what their contribution to participatory democracy in Reykjavik is. This analysis shed new lights on this matter as previous results dated back from 2009 and only showed that 26% of Reykjavik citizens tried to influence their municipality decision making process using online tools (Report 2016).

The results presented in the report had showed that around two third of people living in Reykjavik was familiar or at least heard of Better Neighborhoods and Better Reykjavik. When looking at background of citizens who use these online tools, they present groups of people who are generally more active in terms of political participation and activity, i.e. university-educated, with higher salaries. Namely, the further assessment showed that 43% of citizens with university education had heard about Better Reykjavik compared to only 16% of those who had completed only primary education. Moreover, citizens between 30-60 years are more familiar than residents belonging to the youngest or oldest age groups, which is interesting finding as this option does not particularly appeal to the younger generation often called the computer generation. When it comes to investigating people satisfaction level, this study had included only citizens who had at least heard of these tools, politicians and city officials. 67% of citizens were satisfied with Better Reykjavik compared to 69% of those satisfied with Better Neighborhoods. When it comes to politicians, 47% of them were pleased with Better Reykjavik and 67% with Better Neighborhoods. Finally, around a half of the city officials were happy with both efforts, compared to 40% for Better Reykjavik and 20% for Better Neighborhoods who were not.

**Results and Discussion**

In order to compare the aforementioned tools, we used a framework for systematic analysis and comparison of e-participation platforms proposed by Poplin et al. (2013) called the *participatory cube*. The model is based on previous theories developed by Fung (2006) and Ferber et al. (2007) and incorporates the established ladder of citizen participation proposed by Arnstein (1969). To compare across different e-participation tools, the authors identify three measurements of analysis: the decision power vested in participants, the interactivity of communication, and the provided access to the space of participation. Figure 1 shows that when analyzing our case studies using the participatory cube framework, it reveals little variability in level of interactivity between Better Reykjavik and Better Neighborhood, whereas social media score as a less participatory mode of communication, represented by Facebook closed group of the Breiðholt community (restricted to specific themes and allowing consultation only) and the official Reykjavik Facebook page. Namely, despite of the freedom to participate in social media, dialogue is one-sided and there is no opportunity for consultation or taking part in decisions. One might say that Better Neighborhoods scores the highest on a ladder of participation because it allows the transfer of power from government to citizens, although in a limited area of city budget and only for projects of minor importance.

Based on the analysis, we identified some general recommendations for success across the above dimensions of participation online. With regard to interactive communication, supportive measures should include visualization and use of rich media (e.g. virtual worlds, simulations, audio and video content), real-time interactions (e.g. Q&A sessions, chat, texting, web conferencing), and opportunities for participants' collaboration and self-organization online (e.g. profiles, groups, discussion forums). The space of participation could be expanded through the provision of multilingual and disability-friendly interface, support and alternatives to those with limited access to online platforms, protection of personal data, statistics and tracking for underrepresentation.

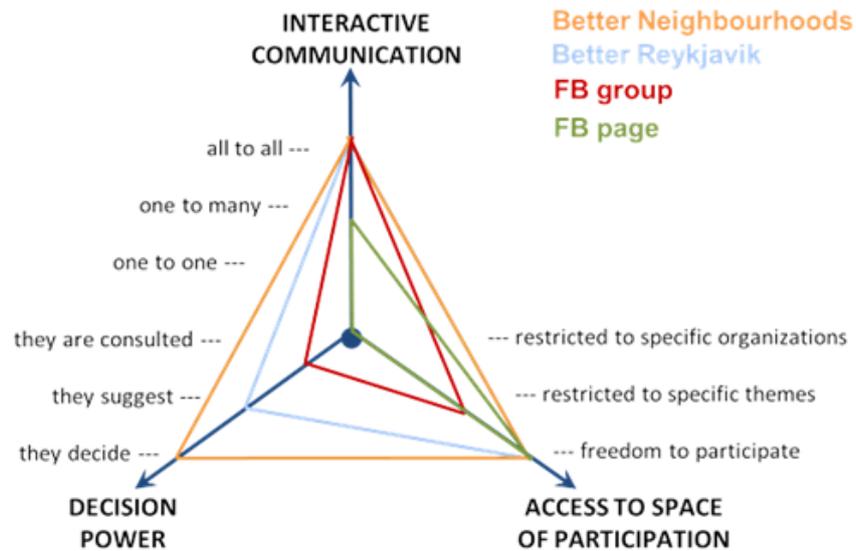

Fig. 1 Participatory cube framework for considered case studies.

Improved outreach could be achieved through social media, online publicity, and local multipliers. Stronger decision power could be vested in participants online by facilitating learning and deliberation (e.g. wikis, web conferencing, and discussion forums) and transparent voting and selection procedures. Last but not least, crucial to the positive outcome of e-participation is the predefinition of participants' impact on final decisions and their engagement in the implementation and evaluation of proposed ideas. With view to the analyzed specific case studies, we iterate the need for mixed use of different media, for example the seamless integration of Facebook communication for promotion of citizens' ideas and voting. One of the few but significant shortcomings of the Better Reykjavik forum is the delayed feedback and lack of transparency when processing and adapting citizens' ideas to legal and technical requirements.

## Conclusions

The benefits of e-participation include the potentials of overcoming time and space constraints, ease of access to information and participation and non-discrimination, and the possibilities of crowdsourcing collective knowledge, and using playfulness to increase participation and attractiveness to young people. However, in terms of empowerment, online deliberation and social media could only supplement ongoing efforts to involve citizens in co-creating their environment and cannot be treated as panacea or substitute to traditional and much necessary forms of democratic governance. Participatory decision-making on matters of public concern justly consumes time and resources, therefore online tools should be applied with consideration of scale and efficiency, i.e. on burning issues for a majority of citizens or small-scale local platforms, and in combination with meetings in real time and space. The budget and workforce allocated to managing online engagement tools should be proportionate to other political and administrative efforts to bring to execution proposed ideas and act on collected feedback in order to satisfy the needs expressed by the communities and not undermine their beliefs about their power to influence decisions.

**Acknowledgment**

The research was supported by the Transport and Urban Development (TUD) COST Action TU1204 (EU Framework Programme Horizon 2020) *People Friendly Cities in a Data Rich World* within the framework of the international winter school *Co-Creating Urban Spaces: The Transformation of the Given City* in Iceland in March 2016. The authors want to recognize the support of Unnur Margrét Arnardóttir and Hilmar Hildar Magnúsarson, Office of the Mayor and Chief Executive Officer, City of Reykjavík; Óskar Dýrmundur Ólafsson, community manager in Breiðholt; and Sjöfn Vilhelmsdóttir, Director of the Institute of Public Administration and Politics, Faculty of Political Science, University of Iceland. The first author would also like to acknowledge support of the National Research Foundation, Prime Minister's Office, Singapore, under its CREATE programme, Singapore-MIT Alliance for Research and Technology Future Urban Mobility IRG and research project "Managing Trust and Coordinating Interactions in Smart Networks of People, Machines and Organizations", funded by the Croatian Science Foundation under the project UIP-11-2013-8813.